\documentstyle[12pt]{article}

\begin{document}

\begin{titlepage}

\begin{flushright}
{\sc TRI-PP-94-20}\\ [.2in]
{\sc \today}\\ [.5in]
\end{flushright}

\begin{center}
{\LARGE
More on direct CP violation in $b \rightarrow d J/\psi$ decays}\\ [.5in]
{\large
Jo\~{a}o M. Soares}\\ [.1in]
{\small
TRIUMF, 4004 Wesbrook Mall, Vancouver, BC Canada V6T 2A3}\\ [.5in]

{\normalsize\bf
Abstract}\\ [.2in]
\end{center}

{\small
Direct CP violation can occur in $B$-meson decays of the type
$b \rightarrow d c \bar{c}$, where the charm-anticharm pair forms a
$J/\psi$. The CP asymmetry requires the contribution to the amplitude
from decays into other states, which rescatter into $d J/\psi$ {\it via} final
state interactions. In particular, states with the same quark content
as $d J/\psi$ can contribute. A perturbative calculation, based on a quark
level description of the rescattering process, gives an asymmetry of about
$1\%$, due to this effect. This makes it the dominant contribution to the
asymmetry, as suggested by an earlier estimate, based on an hadronic picture.}

\end{titlepage}

\section{Introduction}

Direct CP violation in B-meson decays may generate an asymmetry
\begin{equation}
a_{CP} = \frac{\Gamma(B \rightarrow f) -
\Gamma(\bar{B} \rightarrow \bar{f})}{\Gamma(B \rightarrow f) +
\Gamma(\bar{B} \rightarrow \bar{f})},
\label{eq:1}
\end{equation}
between the rates for the CP conjugated processes $B \rightarrow f$ and
$\bar{B} \rightarrow \bar{f}$, even in the absence of $B - \bar{B}$ mixing.
In the standard model, it is predicted that the most significant asymmetries
of this type, which tend to be fairly small
\cite{bss:mechanism,charmless:asym,bdgamma:cp},
occur for Cabibbo suppressed decays. Hence, a large number of $B$-mesons is
necessary, and the observation of the asymmetries in eq.~\ref{eq:1} may be
best achieved at hadronic accelerators. The decays $b \rightarrow q c \bar{c}$
($q = s,d$), where the charm-anticharm pair forms a $J/\psi$, are particularly
suitable for an hadronic machine, given the clean signature from
$J/\psi \rightarrow l^+l^-$. It was pointed out recently by Dunietz
\cite{isi:1}, and later investigated in more detail in ref.~\cite{isi:2},
that a CP asymmetry of type \ref{eq:1} appears in these decays. A rough
estimate suggested that it could be of order $1\%$, in decays of the type
$b \rightarrow d J/\psi$, which is the expected reach of an experiment with a
sample of $10^{10}$ $B$-mesons (the analogous asymmetry in
$b \rightarrow s J/\psi$ is suppressed by a factor of $\sin^2\theta_C$).

The CP asymmetry in $b \rightarrow d J/\psi$ stems from the interference
between the dominant tree amplitude and a small absorptive amplitude that
contributes coherently. That is due to the process
$b \rightarrow i \rightarrow d J/\psi$, where $i$ denotes on-mass-shell
intermediate states with quark content $d u \bar{u}$ or $d c \bar{c}$.
For the former, the process is OZI suppressed, and the rescattering to
$d J/\psi$ occurs at the order $\alpha_s^3$ or {\em via} the electromagnetic
interaction. It generates an asymmetry of the order of {\em a few}
$\times 10^{-3}$ \cite{isi:2}. As for the intermediate states with the same
quark content, $d c \bar{c}$, as the final state, it was first pointed out by
Wolfenstein \cite{wolf:cpt} that they can contribute to the CP asymmetry in
exclusive decays, or in semi-inclusive decays such as $b\rightarrow d J/\psi$
(although, their contribution is absent in the case of the inclusive decay
\cite{charmless:asym}). In ref.~\cite{isi:2}, the nature of this effect was
discussed in terms of the on-mass-shell hadronic states that form the
intermediate state. This approach is reviewed in section 2; it only allows
for a rough estimate which suggests that a contribution to the
CP asymmetry of about $1\%$ is possible. In section 3, I look at this effect
from a different perspective, and try to obtain a better estimate of its
contribution to the asymmetry. In the spirit of quark-hadron duality,
the collection of hadronic intermediate states is replaced by the
corresponding $dc\bar{c}$ quark configuration. The final state
scattering is then treated perturbatively in $\alpha_s$; the absorptive part
of the amplitude and the ensuing CP asymmetry are evaluated at the lowest
order. The results obtained are discussed in section 4.

\section{The hadronic description}

The effect of the final state interactions in a given decay amplitude,
$A_i \equiv A(B \rightarrow i)$, can be described by the S-matrix,
$S = 1 + i T$, for the scattering among different final states.
When the amplitudes in $T$ can be treated perturbatively
\cite{wolf:cpt,wolf:1},
\begin{equation}
A_i  \simeq   A_i^{(0)} +  i \frac{1}{2} \sum_j T_{ij} A_j^{(0)} ,
\label{eq:3}
\end{equation}
where $A_j^{(0)}$ are the weak decay amplitudes in the absence of the final
state interactions. It is the interference between the different terms on the
RHS that generates the CP violating quantity
\begin{eqnarray}
\Delta_i & \equiv & |A_i|^2 - |\bar{A}_i|^2 \nonumber \\
& = & \sum_j \Delta_i^j
\label{eq:4}
\end{eqnarray}
with
\begin{equation}
\Delta_i^j = 2 T_{ij} Im\{ A_j^{(0)\ast} A_i^{(0)}\} .
\label{eq:4a}
\end{equation}
Clearly, $\Delta_i^i = 0$: the rescattering of the final state does not
contribute to the asymmetry, since it does not generate a
term in eq.~\ref{eq:3} with a different CP-odd phase than that of $A_i^{(0)}$.
For the case of the inclusive decay $b \rightarrow dc\bar{c}$,
this means that there is no contribution to the asymmetry
from the intermediate state $dc\bar{c}$. However, the situation is different
for the exclusive or semi-inclusive cases \cite{wolf:cpt}.
In ref.~\cite{isi:2} the decay $B^- \rightarrow J/\psi \pi^-$ was
examined, as an example. Including the absorptive part that is due to the
intermediate states $X = D^0 D^-, D^{\ast -} D^0, J/\psi \rho^-, ... $, that
have the same quark content as the final state $J/\psi \pi^-$, the decay
amplitude is
\begin{eqnarray}
A(B^- \rightarrow J/\psi \pi^-) &=& V_{cb} V_{cd}^\ast T_{\psi\pi^-}
+ V_{tb} V_{td}^\ast P_{\psi\pi^-} \nonumber \\
&+& i \frac{1}{2} \sum_X \left( V_{cb} V_{cd}^\ast T_{X}
+ V_{tb} V_{td}^\ast P_{X} \right) A(X \rightarrow J/\psi \pi^-).
\label{eq:5}
\end{eqnarray}
The weak amplitudes include both tree and penguin contributions, proportional
to $V_{cb} V_{cd}^\ast$ and $V_{tb} V_{td}^\ast$, respectively.
The penguin/tree ratios $P_{\psi\pi^-}/T_{\psi\pi^-}$ and $P_{X}/T_{X}$ will
in general be different, and so the dispersive and absorptive parts of the
amplitude in eq.~\ref{eq:5} will have different CP-odd phases. Then,
the states $X$ will contribute to the CP asymmetry with
\begin{equation}
a_{CP} \simeq Im\{\frac{V_{tb} V_{td}^\ast}{V_{cb} V_{cd}^\ast}\}
 \sum_X \frac{T^{\ast}_{\psi\pi^-} T_X A(X \rightarrow J/\psi \pi^-)}
{|T_{\psi\pi^-}|^2} (\frac{P_X}{T_X} - \frac{P_{\psi\pi^-}}{T_{\psi\pi^-}}).
\label{eq:6}
\end{equation}
There is no reliable way of calculating the scattering amplitudes
$A(X \rightarrow J/\psi \pi^-)$ at the hadronic level (moreover, a large
number of hadronic states, including those with larger multiplicity,
should be included). In ref.~\cite{isi:2}, the asymmetry due to some of the
intermediate states $X$ was estimated, leaving the ratio
\begin{equation}
 \xi_X  \equiv \frac{T^{\ast}_{\psi\pi^-} T_X
A(X \rightarrow J/\psi \pi^-)}  {|T_{\psi\pi^-}|^2}
\label{eq:7}
\end{equation}
as an undetermined parameter. Contributions to the asymmetry of about
\begin{equation}
a_{CP} = \xi_X \times 1\% \times \frac{\eta}{0.4}
\label{eq:8}
\end{equation}
were found ($\eta = - Im\{(V_{tb} V_{td}^\ast)/(V_{cb} V_{cd}^\ast)\}$,
in the Wolfenstein parametrization of the CKM matrix) .

\section{Quark level description}

In view of the difficulties of the approach outlined in the previous section,
an approximate prescription may provide a more complete calculation of the
absorptive amplitude. It amounts to replacing the collection of the hadronic
intermediate states by the corresponding quark configuration $dc\bar{c}$
(in the spirit of the quark-hadron duality). Then, the scattering
to $d J/\psi$ is treated perturbatively in the strong coupling constant,
which is chosen at the $m_b$ scale (for $m_b \simeq 5.0 GeV$ and
$\Lambda_{\overline{MS}}^{(4)} \simeq 200 MeV$, $\alpha_s(m_b) \simeq 0.23$).
Notice that at zeroth order in $\alpha_s$, there are no intermediate states
(other than $d J/\psi$) that contribute to the absorptive amplitude, as
the $J/\psi$ resonance is below the $D-\bar{D}$ threshold, and its overlap
with $\psi^\prime$ can be neglected. At the order $\alpha_s$, the presence
of the gluon removes the kinematical constraint, and the intermediate states
where $c\bar{c}$ appears in a color octet ({\em i. e.} states in the
$c-\bar{c}$ continuum) will contribute. The intermediate states with $c\bar{c}$
in a color singlet will be neglected: they can only contribute at higher orders
in $\alpha_s$, due to the color selection rule, and the weak decay amplitude
into that configuration is color suppressed.

The amplitude for the decay $b \rightarrow d J/\psi$, in the absence of
final state scattering, is
\begin{equation}
A_{d\psi}^{(0)}
= V_{cb} V_{cd}^\ast T_{d\psi} + V_{tb} V_{td}^\ast P_{d\psi} .
\label{eq:9}
\end{equation}
The tree and penguin terms are calculated from the effective Hamiltonian
\begin{eqnarray}
 H_{eff} &=& - \frac{G_{F}}{\sqrt{2}}  \,[ V_{ub} V_{ud}^\ast
( C_{1} {\cal Q}_{1}^u +
C_{2} {\cal Q}_{2}^u ) + V_{cb} V_{cd}^\ast
( C_{1} {\cal Q}_{1}^c +
C_{2} {\cal Q}_{2}^c ) \nonumber \\
& & + V_{tb} V_{td}^\ast \sum_{k=3}^{6} C_{k} {\cal Q}_{k}  + h. c. ]
 \, ,    \label{eq:10}
\end{eqnarray}
where
\begin{eqnarray}
{\cal Q}_{1}^l &=& \bar{d} \gamma^{\mu} (1 - \gamma_5)  b \:\,
\bar{l} \gamma_{\mu}  (1 - \gamma_5) l  \nonumber \\
{\cal Q}_{2}^l &=&  \bar{l} \gamma^{\mu}  (1 - \gamma_5) b \:\,
\bar{d} \gamma_{\mu}  (1 - \gamma_5) l  \nonumber \\
{\cal Q}_{3} &=& \sum_{l=u,d,s,c,b} \bar{d}\gamma^{\mu}  (1 - \gamma_5) b \:\,
\bar{l}\gamma_{\mu}  (1 - \gamma_5) l  \nonumber \\
{\cal Q}_{4} &=& \sum_{l=u,d,s,c,b} \bar{l}\gamma^{\mu}  (1 - \gamma_5) b \:\,
\bar{d}\gamma_{\mu}  (1 - \gamma_5) l  \nonumber \\
{\cal Q}_{5} &=& \sum_{l=u,d,s,c,b}  \bar{d}\gamma^{\mu}   (1 - \gamma_5) b
\:\,
\bar{l}\gamma_{\mu}   (1 + \gamma_5) l  \nonumber \\
{\cal Q}_{6} &=&  - 2 \sum_{l=u,d,s,c,b} \bar{l}   (1 - \gamma_5) b \:\,
\bar{d}  (1 + \gamma_5) l     .   \label{eq:11}
\end{eqnarray}
In the leading-logarithm approximation, the Wilson coefficients at the
scale $m_b$ (and for $\Lambda_{\overline{MS}}^{(4)}$ as above) are
\cite{wilsoncoeff:Buras}
\begin{eqnarray}
  C_1 &=& 0.25 \nonumber \\
  C_2 &=& - 1.11 \nonumber \\
  C_3 &=& 0.011 \nonumber \\
  C_4 &=& -0.026 \nonumber \\
  C_5 &=& 0.008 \nonumber \\
  C_6 &=& -0.032   . \label{eq:12}
\end{eqnarray}
Then,
\begin{equation}
T_{d\psi} =  \frac{G_F}{\sqrt{2}} (C_1 + \frac{1}{N_c} C_2)
m_{\psi} f_{\psi} \epsilon_\mu^\ast \bar{u}_d \gamma^\mu
(1-\gamma_5) u_b ,
\label{eq:13}
\end{equation}
and
\begin{eqnarray}
P_{d\psi} &=&  \frac{G_F}{\sqrt{2}} [C_3 + C_5 + \frac{1}{N_c}
(C_4 + C_6)] \nonumber\\
& & \times m_{\psi} f_{\psi} \epsilon_\mu^\ast \bar{u}_d \gamma^\mu
(1-\gamma_5) u_b .
\label{eq:14}
\end{eqnarray}
The internal momentum of the $c\bar{c}$ pair that forms the
$J/\psi$ is neglected, and the decay constant $f_{\psi}$
is defined by
\begin{equation}
<J/\psi|\bar{c} \gamma_\mu c|0> = m_{\psi} f_{\psi} \epsilon_\mu^\ast .
\label{eq:15}
\end{equation}

The same Hamiltonian gives the amplitude for the decay
$b \rightarrow d (c \bar{c})_8$, where the charm-anticharm pair forms a
color octet,
\begin{equation}
A_8^{(0)}
= V_{cb} V_{cd}^\ast (T_8^V + T_8^A)
+ V_{tb} V_{td}^\ast (P_8^V + P_8^A).
\label{eq:16}
\end{equation}
The superscripts $V$ and $A$ designate the terms that correspond to the
$c\bar{c}$ pair in a vector and in an axial-vector state, respectively.
The latter are given by
\begin{equation}
T_8^A = - \frac{G_F}{\sqrt{2}}  C_2 \frac{1}{2}
\bar{u}_d \gamma^\mu (1-\gamma_5) \lambda^a u_b
\bar{u}_c \gamma_\mu \gamma_5 \lambda^a v_{\bar{c}}
\label{eq:17}
\end{equation}
and
\begin{equation}
P_8^A = - \frac{G_F}{\sqrt{2}} (C_4 - C_6) \frac{1}{2}
\bar{u}_d \gamma^\mu (1-\gamma_5) \lambda^a u_b
\bar{u}_c \gamma_\mu \gamma_5 \lambda^a v_{\bar{c}} .
\label{eq:18}
\end{equation}
As for the former, it is shown below that they do not contribute to the
absorptive part of the $b \rightarrow d J/\psi$ amplitude, if the final
state scattering is treated at the order $\alpha_s$.

The $c\bar{c}$ color octet can scatter to the $J/\psi$ by exchanging a
gluon with the $d$-quark. As pointed out earlier, this gluon exchange is
required not only by the color constraint, but also for kinematical reasons,
since the $J/\psi$ is below the $c\bar{c}$ continuum (a similar effect was
discussed in ref.~\cite{bdgamma:cp} in relation with the CP asymmetry in the
radiative $b$ decays). The convolution of the scattering amplitude
$A(d(c\bar{c})_8 \rightarrow dJ/\psi)$ with $A_8^{(0)}$ gives the contribution
to the absorptive part of the $b \rightarrow d J/\psi$ amplitude:
\begin{equation}
A_{d\psi}^{absorptive} = i \frac{1}{2} \sum \int d\Phi
A(d(c\bar{c})_8 \rightarrow dJ/\psi) A_8^{(0)}  .
\label{eq:19}
\end{equation}
The summation is over the spin and color, and the integral is over the phase
space of the intermediate state quarks. The integration was done analytically,
and the rather cumbersome result is given in the Appendix.

The scattering amplitude $A(d(c\bar{c})_8 \rightarrow dJ/\psi)$ is the sum of
two terms, that correspond to the diagrams with a gluon exchanged between the
$d$-quark and either the $c$- or the $\bar{c}$-quarks. As long
as the internal momentum of the $J/\psi$ is neglected, the corresponding
terms in the expression for $A_{d\psi}^{absorptive}$ are related by charge
conjugation, and they are equal in magnitude.
Because, the $J/\psi$ is a vector state ($C=-1$), the two terms have
the same sign when the $A_8^{(0)}$ amplitude produces
$c\bar{c}$ as an axial vector. Whereas they have the opposite sign, and
they cancel each other, when the $c\bar{c}$ octet forms a
vector (this is nothing else than a manifestation of Furry's theorem).
The amplitude for the decay $b \rightarrow d J/\psi$, with the final state
scattering included to order $\alpha_s$, is then
\begin{eqnarray}
A_{d\psi}
&=& V_{cb} V_{cd}^\ast T_{d\psi} + V_{tb} V_{td}^\ast P_{d\psi}
\nonumber\\
& & + i \frac{1}{2} \sum \int d\Phi A(d(c\bar{c})_8 \rightarrow dJ/\psi)
(V_{cb} V_{cd}^\ast T_8^A + V_{tb} V_{td}^\ast P_8^A) ,
\label{eq:23}
\end{eqnarray}
where the dispersive terms are given in eqs.~\ref{eq:13} and \ref{eq:14},
and the absorptive part is given in the Appendix.
The interference between the terms with relative CP-odd and CP-even phases
in the RHS gives
\begin{eqnarray}
\Delta & \equiv & |A_{d\psi}|^2 - |\bar{A}_{d\psi}|^2
\nonumber \\
& = & 2 Im\{V_{cb} V_{cd}^\ast V_{tb}^\ast V_{td}\} \sum \int d\Phi
 A(d(c\bar{c})_8 \rightarrow dJ/\psi) \nonumber\\
& &  \times (P_8^A T_{d\psi}^{\dag} - T_8^A P_{d\psi}^{\dag}) ,
\label{eq:24}
\end{eqnarray}
where the summation includes the spin and color of the $b$- and $d$-quarks,
and the $J/\psi$ polarization. It follows that
\begin{eqnarray}
\Delta
& = & - Im\{V_{cb} V_{cd}^\ast V_{tb}^\ast V_{td}\}
[(C_4-C_6)(C_1+\frac{1}{N_c}C_2) \nonumber\\
& & - C_2 (C_3 +C_5 +
\frac{1}{N_c} (C_4 + C_6))] \nonumber\\
& &  \times  G_F^2 \alpha_s m_b^3 m_{\psi} f_{\psi}^2 \frac{32}{3}
\frac{(1-z)^2 (z^2-3)}{\sqrt{z} (2-z)^2} ,
\label{eq:25}
\end{eqnarray}
with $z=(m_{\psi}/m_b)^2$. This gives the CP asymmetry, in the semi-inclusive
decay $b \rightarrow d J/\Psi$,
\begin{eqnarray}
a_{CP} & \simeq & \frac{|A_{d\psi}|^2 - |\bar{A}_{d\psi}|^2}
{2 |V_{cb} V_{cd}^\ast T_{d\psi}|^2} \nonumber \\
& = & Im\{ \frac{V_{tb} V_{td}^\ast}{V_{cb} V_{cd}^\ast} \}
\frac{(C_4-C_6)(C_1+\frac{1}{N_c}C_2) - C_2 (C_3 +C_5 +
\frac{1}{N_c} (C_4 + C_6))}{(C_1+\frac{1}{N_c}C_2)^2} \nonumber\\
& & \times \alpha_s \frac{8}{9} \frac{(1-z) (z^2-3)}{(1+2z) (2-z)^2} .
\label{eq:26}
\end{eqnarray}
Following the usual prescription of setting $N_c = \infty$
(so that the strength of the color suppression in
$\Gamma(b \rightarrow d J/\psi)$ is in good agreement with what is measured
in the analogous decays of the type $b \rightarrow s J/\psi$)
\cite{colorsuppr:disc},
\begin{equation}
a_{CP} = 1.1\% \times \frac{\eta}{0.4} .
\label{eq:27}
\end{equation}

\section{Conclusion}

The value of the CP asymmetry in eq.~\ref{eq:27} confirms the earlier
suspicion \cite{isi:2} that the contribution from the
$dc\bar{c}$ intermediate states may be important.
Indeed, if the assumptions in which this calculation is based are correct
(namely, the use of a quark configuration for the intermediate
state, and an expansion in $\alpha_s$ for the final state scattering),
the absorptive part of the $b \rightarrow d J/\psi$ amplitude is
dominated by the rescattering from states that contain a $c\bar{c}$ pair
in a color octet, {\em i. e.} states in the continuum above the $D-\bar{D}$
threshold. Their contribution to the CP asymmetry is somewhat lowered by
the fact that it must be proportional to a ratio of penguin to tree amplitudes
(both are necessary in order to generate a relative CP-odd phase). Still, it
dominates over the contribution from the OZI suppressed process
$b \rightarrow du\bar{u} \rightarrow d J/\psi$ \cite{isi:2,isi:3}.

An important source of uncertainty is the very bothersome fact that,
at present, the strength of the color suppression in decays such
as $b \rightarrow q J/\psi$ ($q=s$ or $d$) is not well understood.
The prescription of dropping all non-leading terms in $1/N_c$, that
I adopted in here, allows to reproduce the branching ratios that have been
measured, but it is not well founded theoretically, nor confirmed by data
from other types of $B$ decays \cite{colorsuppr:disc}.
For the moment, it provides a systematic
framework to derive quantitative predictions. However, it is quite possible
that some new mechanism is at work that would dominate the decay rate, and
most likely affect the value of the CP asymmetry.
Hence the interest in pursuing an experimental search, given the potential
of present and future facilities for probing the asymmetry close to the level
predicted in here.

\section*{Acknowledgements}

I wish to thank Lincoln Wolfenstein, Isi Dunietz and Per Ernstrom
for enlightening discussions and helpful criticism. This work was
partly supported by the Natural Science and Engineering Research
Council of Canada.

\section*{Appendix}

The scattering amplitude $A(d(c\bar{c})_8\rightarrow dJ/\psi)$ is the sum
of a term
\begin{eqnarray}
\bar{t} &=& \alpha_s \pi
\bar{u}_d \gamma^\mu \lambda^a u_d^\prime
\bar{v}_{\bar{c}}^\prime \gamma^\mu \lambda^a v_{\bar{c}}
\frac{1}{(p_d^\prime - p_d)^2}
\label{eq:A1}
\end{eqnarray}
(the quantities $u_d^\prime$, $\bar{v}_{\bar{c}}^\prime$, $p_d^\prime$, and
later $p_{\bar{c}}^\prime$, correspond to the intermediate state quarks),
that corresponds to the gluon exchange between the $d$- and the
$\bar{c}$-quarks, and an analogous term $t$ due to the gluon exchange
between the $d$- and the $c$-quarks. In the expression for the absorptive
amplitude in eq.~\ref{eq:19}, the contributions from $t$ and $\bar{t}$ are
related by charge conjugation, and it follows that
\begin{eqnarray}
A_{d\psi}^{absorptive} &= &  i \sum \int d\Phi
\bar{t} (V_{cb} V_{cd}^\ast T_8^A + V_{tb} V_{td}^\ast P_8^A) .
\label{eq:A2}
\end{eqnarray}
The tree and penguin terms, $T_8^A$ and $P_8^A$, are given in eqs.~\ref{eq:17}
and \ref{eq:18}, and
\begin{equation}
d\Phi \equiv (2\pi)^4 \delta^4
(p_{\bar{c}} + p_d - p_{\bar{c}}^\prime - p_d^\prime)
\frac{d^3 p_{\bar{c}}^\prime}{(2\pi)^3 2 p_{\bar{c}}^{\prime 0}}
\frac{d^3 p_d^\prime}{(2\pi)^3 2 p_d^{\prime 0}} .
\label{eq:A3}
\end{equation}
The hadronization of the $c\bar{c}$ pair, that forms the $J/\psi$
in the final state, is described by a single parameter:
the magnitude of the $J/\psi$ wavefunction at the origin or, equivalently,
the decay constant defined in eq.~\ref{eq:15}. The relations
\begin{eqnarray}
<J/\psi|\bar{c} \sigma_{\mu\nu} c|0> &=& i f_{\psi}
(p_{\psi\mu} \epsilon_\nu^\ast - p_{\psi\nu} \epsilon_\mu^\ast) \nonumber\\
<J/\psi|\bar{c} \sigma_{\mu\nu} \gamma_5 c|0> &=& \frac{1}{2} i
\epsilon_{\mu\nu\alpha\beta} <J/\psi|\bar{c} \sigma^{\alpha\beta} c|0>
\label{eq:A4}
\end{eqnarray}
are also useful.
Summing over the spin and color of the intermediate state quarks, and
integrating over their phase space, I obtain the following result
\begin{eqnarray}
A_{d\psi}^{absorptive}
&=& - i \frac{G_F}{2\sqrt{2}} \alpha_s f_{\psi}
[V_{cb} V_{cd}^\ast C_2 + V_{tb} V_{td}^\ast (C_4 - C_6)] \frac{8}{9} \Omega
\label{eq:A5}
\end{eqnarray}
where
\begin{eqnarray}
\Omega &=& m_{\psi} \epsilon^{\sigma\ast}
\{ \bar{u}_d \gamma_\sigma (1-\gamma_5) u_b
[ ({\cal C}_1 + {\cal C}_2)  \frac{p_b.p_d}{m_{\psi}^2}
+ {\cal D} + 3 {\cal G} ]
\nonumber \\
& &- \bar{u}_d (1+\gamma_5) u_b ({\cal C}_1 + {\cal C}_2)
\frac{m_b p_{b\sigma}}{2 m_{\psi}^2}
\nonumber \\
& &-  \bar{u}_d \gamma_\nu \gamma_\sigma \gamma_\mu (1-\gamma_5) u_b
({\cal C}_1 - {\cal C}_2) i \epsilon^{\alpha\mu\beta\nu}
\frac{p_{b\alpha} p_{d\beta}}{4 m_{\psi}^2} \}
\nonumber \\
& &+ \frac{1}{4}
(p_\psi^\sigma \epsilon^{\delta\ast} - p_\psi^\delta \epsilon^{\sigma\ast})
\{ - \bar{u}_d \gamma_\sigma (1+\gamma_5) u_b
\frac{m_b}{m_{\psi}^2} [ ({\cal C}_1 + {\cal C}_2) p_{d\delta}
+ 2 {\cal D} p_{\psi\delta}]
\nonumber \\
& & - \bar{u}_d (1-\gamma_5) u_b 2 ({\cal C}_1 + {\cal C}_2 - 2 {\cal D})
\frac{p_{b\delta} p_{d\sigma}}{m_{\psi}^2}
\nonumber \\
& &+ \bar{u}_d \gamma_\sigma (1-\gamma_5) u_b \frac{2}{m_{\psi}}
[({\cal B}- {\cal A}) p_{d\delta} + {\cal A} p_{b\delta}]
\nonumber \\
& &- \bar{u}_d \gamma_\sigma \gamma_\delta (1-\gamma_5) u_b
[ ({\cal C}_1 + {\cal C}_2) \frac{p_b.p_d}{2 m_{\psi}^2}
+ \frac{1}{2} {\cal D} + 4 {\cal G} ]
\nonumber \\
& &+  \bar{u}_d \gamma_\nu \gamma_\sigma \gamma_\delta \gamma_\mu
(1-\gamma_5) u_b ({\cal C}_1 - {\cal C}_2) i
\epsilon^{\alpha\mu\beta\nu} \frac{p_{b\alpha} p_{d\beta}}{4 m_{\psi}^2} \}
\nonumber \\
& &+ \frac{1}{4} i \epsilon^{\sigma\delta\lambda\rho}
p_{\psi\lambda} \epsilon^\ast_\rho
\{ \bar{u}_d \gamma_\sigma (1+\gamma_5) u_b
\frac{m_b}{m_{\psi}^2}
[ ({\cal C}_1 + {\cal C}_2) p_{d\delta} + 2 {\cal D} p_{\psi\delta}]
\nonumber \\
& &+ \bar{u}_d (1-\gamma_5) u_b 2 ({\cal C}_1 + {\cal C}_2 - 2{\cal D})
\frac{p_{b\delta} p_{d\sigma}}{m_{\psi}^2}
\nonumber \\
& &+ \bar{u}_d \gamma_\sigma (1-\gamma_5) u_b \frac{2}{m_{\psi}}
[({\cal B}- {\cal A}) p_{d\delta} - {\cal A} p_{b\delta}]
\nonumber \\
& &- \bar{u}_d \gamma_\sigma \gamma_\delta (1-\gamma_5) u_b
[ ({\cal C}_1 + {\cal C}_2) \frac{3 p_b.p_d}{2 m_{\psi}^2}
+ \frac{3}{2} {\cal D} + 4 {\cal G} ]
\nonumber \\
& &+ \bar{u}_d \gamma_\sigma \gamma_\delta (1+\gamma_5) u_b  {\cal A}
\frac{m_b}{m_{\psi}}
\nonumber \\
& &-  \bar{u}_d \gamma_\nu \gamma_\sigma \gamma_\delta \gamma_\mu
(1-\gamma_5) u_b ({\cal C}_1 - {\cal C}_2) i
\epsilon^{\alpha\mu\beta\nu} \frac{p_{b\alpha} p_{d\beta}}{4 m_{\psi}^2} \} .
\label{eq:A6}
\end{eqnarray}
The quantities ${\cal A}$,  ${\cal B}$,  ${\cal C}_1$,  ${\cal C}_2$,
${\cal D}$, ${\cal F}$, and ${\cal G}$, are defined by
\begin{eqnarray}
4\pi m_{\psi}^2 \int d\Phi \frac{1}{(p_d^\prime - p_d)^2}
p^\prime_{d\alpha} &=& {\cal A} p_{\psi\alpha} + {\cal B} p_{d\alpha}
\nonumber\\
8\pi m_{\psi}^2 \int d\Phi \frac{1}{(p_d^\prime - p_d)^2}
p^\prime_{d\alpha} p^\prime_{\bar{c}\beta} &=&
{\cal C}_1 p_{\psi\alpha}p_{d\beta} + {\cal C}_2 p_{\psi\beta}p_{d\alpha}
+ {\cal D} p_{\psi\alpha} p_{\psi\beta} \nonumber \\
& & + {\cal F} p_{d\alpha} p_{d\beta} + {\cal G} m_{\psi}^2 g_{\alpha\beta} .
\label{eq:A7}
\end{eqnarray}
Performing the integrals, it follows that
\begin{eqnarray}
{\cal A} &=& - \frac{z}{2-z} \nonumber\\
{\cal B} &=& I + \frac{2 z}{(1-z)(2-z)} \nonumber\\
{\cal C}_1 &=& - \left(\frac{z}{2-z}\right)^2 \nonumber\\
{\cal C}_2 &=& I + \frac{2z}{1-z} -
\left(\frac{z}{2-z}\right)^2  \nonumber\\
{\cal D} &=& - \frac{z}{(2-z)^2} \nonumber\\
{\cal G} &=& - \frac{1}{2} \left( \frac{1-z}{2-z} \right)
\label{eq:A8}
\end{eqnarray}
($z=(m_{\psi}/m_b)^2$). The divergent integral
\begin{equation}
I = - \frac{z}{1-z} \int_{-1}^{+1} dx \frac{1}{1-x}
\label{eq:A9}
\end{equation}
corresponds to a mass singularity (due to taking $m_s=0$ or $m_c=m_{\psi}/2$).
However, the dependence on $I$ cancels in the expression for $\Delta$,
and so the CP asymmetry is free of mass singularities (and of IR divergences).


\begin{thebibliography}{Abcd:efgh}
\bibitem{isi:3}
The contribution from the $du\bar{u}$ intermediate state,
with $u\bar{u}$ in a color octet, was not included in ref.~\cite{isi:2}.
Unlike the color singlet case, there is no 1-photon final state scattering
and only the contribution of order $\alpha_s^3$ remains.
\end{thebibliography}
\end{document}